 \documentclass[final,5p,times,twocolumn]{elsarticle}

\usepackage{graphicx}
\usepackage{float}
\usepackage{amssymb}
\usepackage{graphicx}
\usepackage{bm}
\usepackage{times}
\usepackage{hyperref}

\begin{document}

\begin{frontmatter}


\title{Traveling Wave MRI at 21.1 T: Propagation below Cut-off for Ultrahigh Field Vertical Bore System}

\author[label1]{Alexey A. Tonyushkin\corref{cor1}} 
\ead{atonyushkin@mgh.harvard.edu}
\author[label3,label4]{Jose A. Muniz} 
\author[label3,label4]{Samuel C. Grant} 
\author[label1,label2]{Andrew JM Kiruluta}

\cortext[cor1]{Corresponding author}
\address[label1]{Radiology Department, Massachusetts General Hospital, Harvard Medical School, 55 Fruit St, Boston, MA 02114, US}
\address[label3]{National High Magnetic Field Laboratory, 1800 East Paul Dirac Drive, Tallahassee, FL 32310, US}
\address[label4]{Chemical and Biomedical Engineering, The Florida State University, 2525 Pottsdamer St, Tallahassee, FL 32310, US}
\address[label2]{Physics Department, Harvard University, 17 Oxford St, Cambridge, MA 02138, US}

\begin{abstract}
At high magnetic field strengths ( $> 4$ T), the propagation wave vector of the excitation field $B_1$ can no longer be ignored as the wavelength becomes commensurate or smaller than the imaging field of view (FOV), particularly for high dielectric media. The emergence of this propagation wave vector renders the excitation field variant in both time and space. In this paper, the propagation of RF excitation waves in a high field vertical NMR system is demonstrated for the first time with an appropriate screened dielectric waveguide excited by a simple loop coil to enable below cutoff propagation for imaging in the far field. Uniquely, aqueous samples provide the high permittivity required to modify the cut-off requirements of this transmission system. Theory and simulations corroborate the observed propagating modes field patterns. The ease of construction and implementation of this setup permits use in a variety of high field systems. At a minimum, this design provides a larger FOV that is limited primarily by the three-dimensional linearity of the gradient coil.  More critically, it also opens up the use of multimode excitation for selective nulling or maximum excitation of the FOV along the direction of wave propagation.
\end{abstract}
\begin{keyword}
traveling wave MRI \sep ultra-high field \sep 21 T NMR \sep waveguide modes
\end{keyword}
\end{frontmatter}

\section{Introduction}
\label{Intro}
Owing to the low wavelengths in tissues, RF waves at ultra-high fields can generate complex phenomena commonly observed in optics but traditionally negligible in NMR experiments, including phase modulation of the excitation field as well as wave interference and diffraction. The above phenomena are also naturally associated with the propagation effects in media~\cite{Kiruluta2006,Kiruluta2007} that offer several advantages to RF manipulations. Specifically, propagating or traveling wave (TW) can be the most effective way to deliver RF power at ultra-high fields to a large region of interest (ROI) and it can naturally provide variety of modes~\cite{MRM23107}. The propagation of waves inside an electrodynamic system with cylindrical geometry of an MR scanner may be implemented by using a hollow metal waveguide, either by using the bore of the scanner~\cite{Brunner2009}, its shield or a specially constructed and dedicated waveguide~\cite{VazquezAPJ13}. 

Originally implemented in large-bore clinical scanners at 7.0 T~\cite{BrunnerISMRM08}, the TW technique is not easily accessible at higher fields ( $>$ 7 T) since the typical diameters of bores of pre-clinical animal and vertical magnets are generally small compared to the free-space wavelength. Therefore, such waveguides are below the cut-off requirements for wave propagation. Even at 7 T, only the lowest TE mode of cylindrical waveguide can propagate in a hollow bore due to stringent cut-off wavelength requirements. Higher modes (TE, TM and hybrid modes) of a cylindrical waveguide can be propagated only through the use of high permittivity dielectrics~\cite{TonyISMRM11,BrunnerMRM2011}. However, uncontrolled propagating mode excitation in dielectrics may result in non-uniform $B_1$ field distribution, as was early evidenced at high field MRI in the context of dielectric resonance~\cite{Kangarlu1999}. 
Alternatively, wave propagation inside a scanner can be facilitated by a coaxial transmission line~\cite{AltMRM12,CMR20226,MRM24496,MRM24807}.

In this work, a screened dielectric waveguide was employed to demonstrate TW MR imaging below cut-off in a vertical 21.1-T (900-MHz) ultra-wide-bore (105-mm) NMR magnet, the highest field currently available for imaging~\cite{FuGrant2005}. In particular, we implemented an RF transmission line that consists of a cylindrical waveguide partially filled with a concentric dielectric sample, which effectively modifies the cut-off requirements in the bore to achieve RF mode propagation within a relatively small bore. Underscoring its ease in implementation, this waveguide system was mounted with only minor modifications to an existing homebuilt probe frame used for in vivo animal imaging. As such, this design could be implemented in almost any vertical or pre-clinical system for either imaging or spectroscopy applications with minimal effort. The main advantage of TW approach is the removal of the spatial limitations of a near-field RF coil to increase the usable FOV up to the linearity limitations of an imaging gradient or the physical sample constraints of a spectroscopic experiment. Thus, such system would allow large throughput in imaging small animals with a potential of high-speed imaging. 

\section{Methods}
\subsection{Theory}
\label{theory}
Hollow metal waveguides are widely used for microwave transmission in radar applications. The main advantages of such waveguides are ease of specific mode excitation, low power loss for dominant modes and robust implementation. With current cryogenic technology, MR scanners naturally provide a cylindrical geometry with metal boundaries that are naturally formed by the magnet bore, shim coils, gradient coils or RF shields. At ultra-high magnetic fields and the equivalently high Larmor frequency ($\omega$) for $^1$H imaging, the condition for waveguide mode propagation, known as cut-off, can be fulfilled provided that the bore is sufficiently large. In a hollow waveguide, there exist two basic types of modes: transverse electric (TE) and transverse magnetic (TM). Furthermore, the magnitude of each component of the fields (B and E) has spatial dependence, so the propagating modes in a waveguide in cylindrical coordinates $(r,  \phi, z)$ can be expressed as
$$
{\bf E}({\bf r})={\bf E}(r,\phi)e^{-i(hz-\omega t)} ,
{\bf B}({\bf r})={\bf B}(r,\phi)e^{-i(hz-\omega t)} .
$$
The transverse spatial dependence of the modes is defined by the Helmholtz equation and characterized by the integer numbers {\em m, n} that are related to azimuthal and radial field variations, respectively. 
The propagating vector {\em h} of the TE mode is given by the dispersion relationship~\cite{Jackson1998}
\begin{equation}
\label{h}
 h_{mn}=\sqrt{k^2-(\chi'_{mn}/a)^2} .
\end{equation}
Here, {\em k} is a k-vector of free propagating wave inside a medium, {\em a} is the radius of the guide, $\chi'_{mn}$ is the {\em n}-th zero of the derivative of the Bessel function $J_m$ of the first kind of order {\em m} (m=0, 1, 2É). Setting Eq.~(\ref{h}) to be a real number, gives the lowest (cut-off) frequency for the TE mode  
 $$
 f_c=\chi' _{mn}/(2\pi a\sqrt{\mu \varepsilon}) .
 $$ 
The critical wavelength of a mode is $\lambda_c=c/f_c$. The complete solutions for B-field components (here we follow MRI notation and use {\bf B} field instead of {\bf H} field, which is equivalent for $\mu$=1) for TE waves inside a waveguide are given by analytical expressions below~\cite{Jackson1998}:
\begin{eqnarray}
\label{TEfield}
 \nonumber 
B_z=c_m J_m \left( \chi_{mn}\, r/a \right) \cos (m\phi) e^{-ih_{mn} z},\\ 
B_r=\frac{-i a h_{mn} c_m}{\chi^2_{mn}} J'_m \left( \chi_{mn}\, r/a \right) \cos (m\phi) e^{-ih_{mn} z} , \\ \nonumber
B_{\phi}=\frac{i m a^2 h_{mn} c_m}{\chi^2_{mn} r} J_m \left( \chi_{mn}\, r/a \right) \sin (m\phi) e^{-ih_{mn} z}.
\end{eqnarray}
The respective E-field components can be derived from the impedance condition: ${\bf B}=[\hat{z} \times {\bf E} ]/\eta_{TE}$, where $\eta$ is the transverse impedance of the mode. Here, $B_z$ component does not contribute to $B_1$ field, but it is important to consider for the specific mode excitation method. The field for TM modes can be derived from the Helmholtz equation by setting $B_z=0$ and the dispersion relationship is modified by substituting $\chi'_{mn}$  for $\chi_{mn}$ -- the n-th zero of the Bessel function.

Specifically for MR applications, the lowest modes with the least number of nodes in the $B_1$ are preferred to maintain field homogeneity. Such modes are: TE$_{11}$, TM$_{01}$, TE$_{21}$, TE$_{01}$ and TM$_{11}$. Though on first blush preference for TM modes might be considered because of the lack of a $B_z$ component, power considerations could play an important role in excitation of TM modes versus the generally more efficient TE modes with fewer nodes in the $B_1$. The higher index modes can also be useful for a SENSE method as proposed 
by~\cite{BrunnerMRM2011,Pang2011}.

The main challenge for the traveling wave approach in most conventional MR scanners and NMR spectrometers is the relatively small bore to critical wavelength ratio $(a/\lambda_c)$ even at ultra-high fields. In a 7-T clinical MR system with a bore of $\sim$60 cm and $^1$H Larmor frequency of 298 MHz, which corresponds to a free space wavelength of 100.6 cm, a single TE$_{11}$ can propagate inside the waveguide without a dielectric. To date, this is the only MR system with a sufficient $(a/\lambda_c)$ ratio for wave propagation in a hollow scanner bore. However, since $\lambda \sim 1/\sqrt{\varepsilon_r}$,
the critical wavelength ratio criteria can be met in a smaller bore magnet using a high permittivity dielectric inside the bore. For the highest magnetic field available for MRI, namely the 21.1-T, 10.5-cm magnet employed in this study, the $^1$H Larmor frequency of 900 MHz (33.3-cm free space wavelength) produces a maximum critical wavelength of 18 cm for the TE$_{11}$ mode.  With a uniform dielectric having a $\varepsilon_r > 4$ inside the bore there is at least one propagating mode supported by such a system. For a non-conductive, lossless dielectric with $\varepsilon_r =80 $, a typical number for most water-based phantoms, the traveling wave would have a spatial period of 3.8 cm along z-axis. Recently, \cite{BrunnerMRM2011} demonstrated how several dielectric rods with high permittivities can facilitate multiple mode propagation inside a 7-T human system, while pointing out that a uniform dielectric-filled bore would be impractical in a clinical system. Fortunately, for ultra-high field NMR spectrometers and microimaging systems with bore sizes below approximately 10 cm, a single dielectric rod can nearly uniformly fill the bore to facilitate a number of propagating modes. 

Due to the non-zero gap (cladding) between the metal boundary and dielectric insert, the field pattern of Eq.~(\ref{TEfield}) needs to be modified to allow hybrid modes EH with $E_z, B_z \neq 0$. Such a screened dielectric waveguide partially filled with a concentric dielectric rod represents a medium with radially anisotropic $\varepsilon_r$, and therefore, generally is not solved analytically~\cite{TonyISMRM11}. However, for pure TE modes and in many practical cases, where the cladding or air gap between the dielectric rod and the metal screen is small compared to the wavelength (high fill factor), the $B_1$ map of modes are similar to the ones of uniform cylindrical waveguide. The hybrid modes exist in the range of frequencies $f >f'_c$  with modified critical frequency $f'_c$, where the dominant mode is HE$_{11}$ followed by TE$_{01}$, TM$_{01}$, HE$_{21}$, etc. While numerical simulations produce hybrid modes, for qualitative purpose we imply pure TE(TM) modes where applicable. The subtle difference between the uniform dielectric waveguide and partially filled guide is in the less stringent cut-off requirement in the latter case and therefore larger variety of modes present in such guide. As a result the actual propagating constants (and the periods) of the modes differ from the analytical values calculated from the expression for $h_{mn}$ and have to be numerically calculated. 

In addition to the small bore of the MR spectrometers, there is a relatively long distance ($\sim 1.5$ m for the 21.1-T magnet) from the bore opening to the center of field. Therefore, TW method has to rely on an efficient excitation of the propagating modes inside the magnet bore~\cite{AndreychenkoISMRM09}. Since the free-space wavelength is much larger than the bore, $ka \ll 1$, the patch antenna, which is commonly used at 7 T TW MRI~\cite{ZhangISMRM09,VanBergKlompISMRM09,KroezeISMRM10}, is not practical in animal scanners. Therefore, we use an RF loop-coil as an excitation probe to feed into the cylindrical waveguide~\cite{Webb2010,WigginsISMRM10}. The loop-coil is used both for excitation and reception. This coil can be made small enough to be inserted within the magnet bore to reduce the length of the dielectric and ensure that all components of the RF transmission line are within the magnet bore. Furthermore, because it is coupling into the waveguide structure and not acting as an imaging coil itself, the loop-coil can be aligned at different orientations to the main magnetic field $B_0$ to induce proper selection of the dominant propagating mode. 

The near field approximation in spherical coordinates ($r, \theta, \phi$) for a loop probe with a constant current $I_0$ is given by
~\cite{Balanis2005}
\begin{eqnarray}
\label{eq2}
\nonumber 
B_r=I_0 \cos \theta \frac{a^2}{r^3} \\
B_\theta=I_0 \sin \theta \frac{a^2}{4r^3} \\ \nonumber
E_\phi=-i\eta I_0 \sin \theta \frac{(ka)^2}{4kr^2} .
\end{eqnarray}
The exact solution for the mode excitation in a waveguide can be obtained from the Maxwell equations knowing the distribution of the electric and magnetic dipoles. However, by knowing the $E$ (for electric dipole) and $B$ (for magnetic dipole) fields extrema of the specific mode it is possible to predict its efficient excitation for certain cases. If loop placed in parallel with $B_0$ (concentric with the waveguide) then $\theta =0$ and the magnetic field $B=B_r \| B_0$. In conventional MRI such field would not excite spins (with exception of the smaller peripheral inhomogenious field), however, this magnetic field couples into one of TE mode through maximum overlap (mode weight) with the leading $B_z$ component in Eq.(\ref{TEfield}). The mode index {\em m} depends on the position of the coil with respect to the center of the guide so, for example, in the axial symmetric case we expect excitation of TE$_{01}$ mode. It is known~\cite{Jackson1998} that this mode is a special mode for a cylindrical waveguide as it has the lowest losses among all the modes and therefore is the most efficient way to deliver RF energy to the remote imaging volume. In addition, TE$_{01}$ mode is degenerate to TM$_{11}$ mode with respect to propagating constant (period of the wave in the waveguide), thus one mode can be locally converted into the other due to presence of inhomogeneous boundaries such as additional dielectric. 
Similar analysis can be applied to the the orthogonal orientation of the loop when $\theta =90$. In such case, $B=B_\theta \| B_1$ and the probe preferentially excites lowest TE$_{11}$ mode through $B_r, B_\phi$ components in Eq.(\ref{TEfield}). In addition, various TM (hybrid) modes can be excited depending on the coil position, relative size, and dielectric constant.

\subsection{Simulations}
\label{simu}
To validate theoretical predictions we carried out calculations of the loop-coil matching and coupling into the screened dielectric guide using finite differences time domain software Microwave Studio (CST, Darmstadt, Germany). 
\begin{figure}[h!]
\begin{center}
\includegraphics[width=3.2 in]{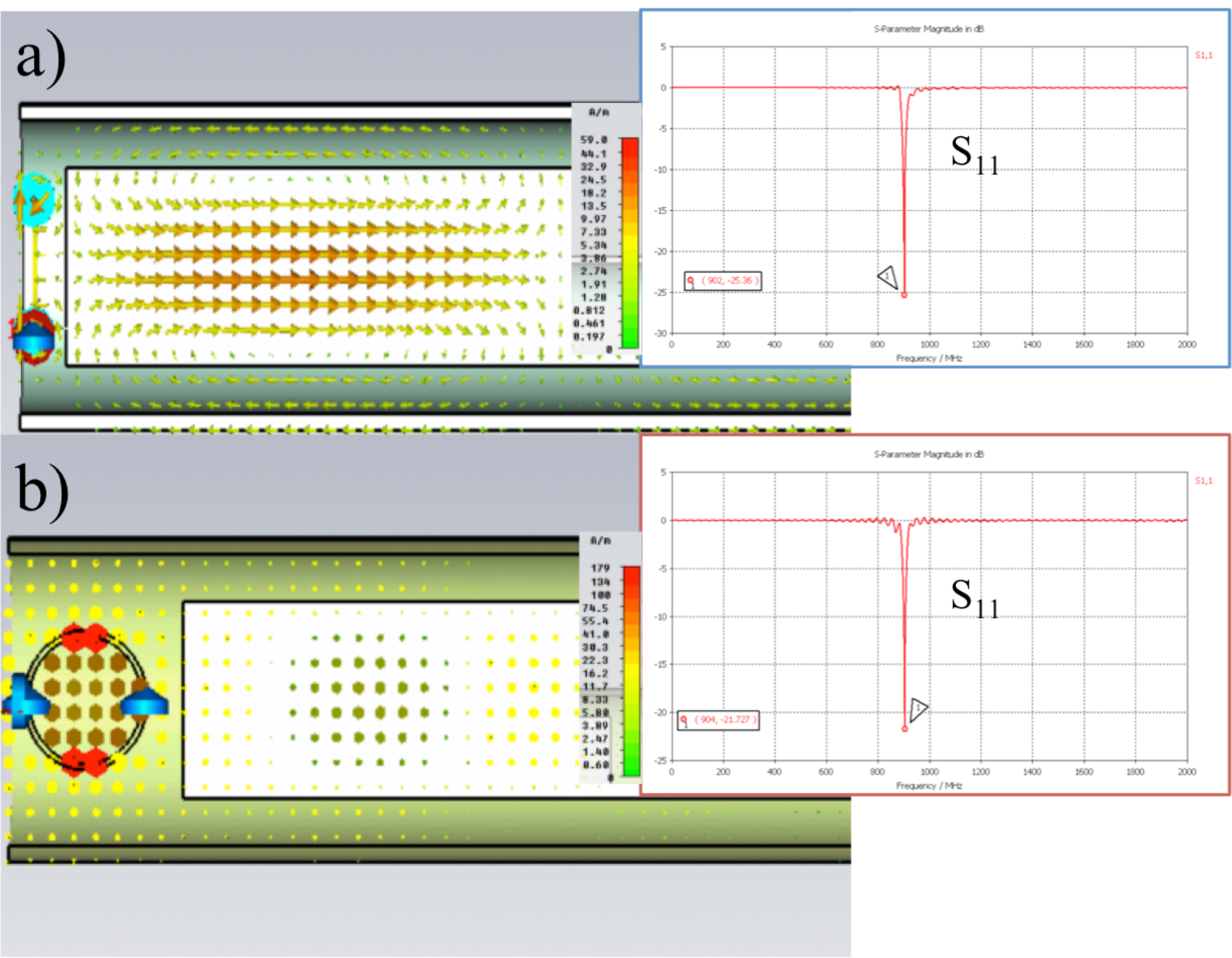}
\end{center}
\caption{ Magnetic field vector map of mode coupling into a partially filled dielectric waveguide for a) parallel, b) orthogonal coil geometries. The inserts show simulated $S_{11}$ parameter with peaks at: (902 MHz, -25 dB) and (904 MHz, -21 dB) for parallel and orthogonal cases, respectively.}
\label{cst}
\end{figure}
Figure~\ref{cst} show vector plot of the magnetic field map for both coil orientation placed at the entrance of the dielectric tube filled with DI water. The resulting mode patterns in the guide correspond to TE$_{01}$ and TE$_{11}$ for two orientations and agree with qualitative conclusions on mode coupling in Sec.~\ref{theory}. The efficiencies of the modes coupling into the waveguide can be characterized by simulating S$_{11}$ parameter that are presented in the insets in Fig.~\ref{cst}. The simulations produced peaks in S$_{11}$ at (902 MHz, -25 dB) for the TE$_{01}$ mode and (904 MHz, -21 dB) for the TE$_{11}$ mode. To increase the coupling efficiencies the coils need to be placed in tight contact with the dielectric.

We further simulated modes in frequency domain by using an eigenmode solver in Comsol Multiphysics (Burlington, MA). To eliminate a trivial coaxial cut-off free waveguide case, for which most of the field is concentrated outside the dielectric, only a high-permittivity, low loss dielectric rod of deionized water placed concentrically within the cylindrical waveguide. 
\begin{figure}[tb!]
\begin{center}
\includegraphics[width=3.2 in]{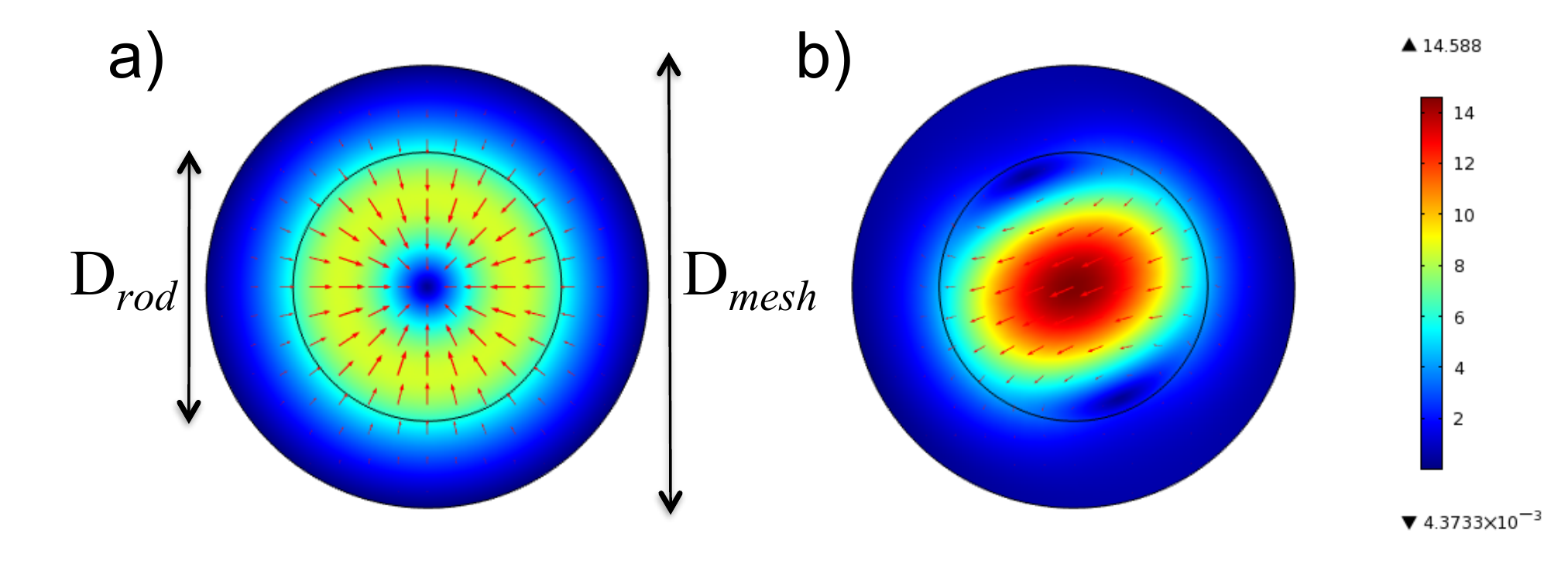}
\end{center}
\caption{ $B_1$ map of two dominated modes in a partially filled dielectric waveguide: a) TE$_{01}$ b) TE$_{11}$; the arrows represent $B_1$ field (here and elsewhere color bar has a relative scale). }
\label{modes}
\end{figure}
%

\begin{figure}[t!]
\begin{center}
\includegraphics[width=3.6 in]{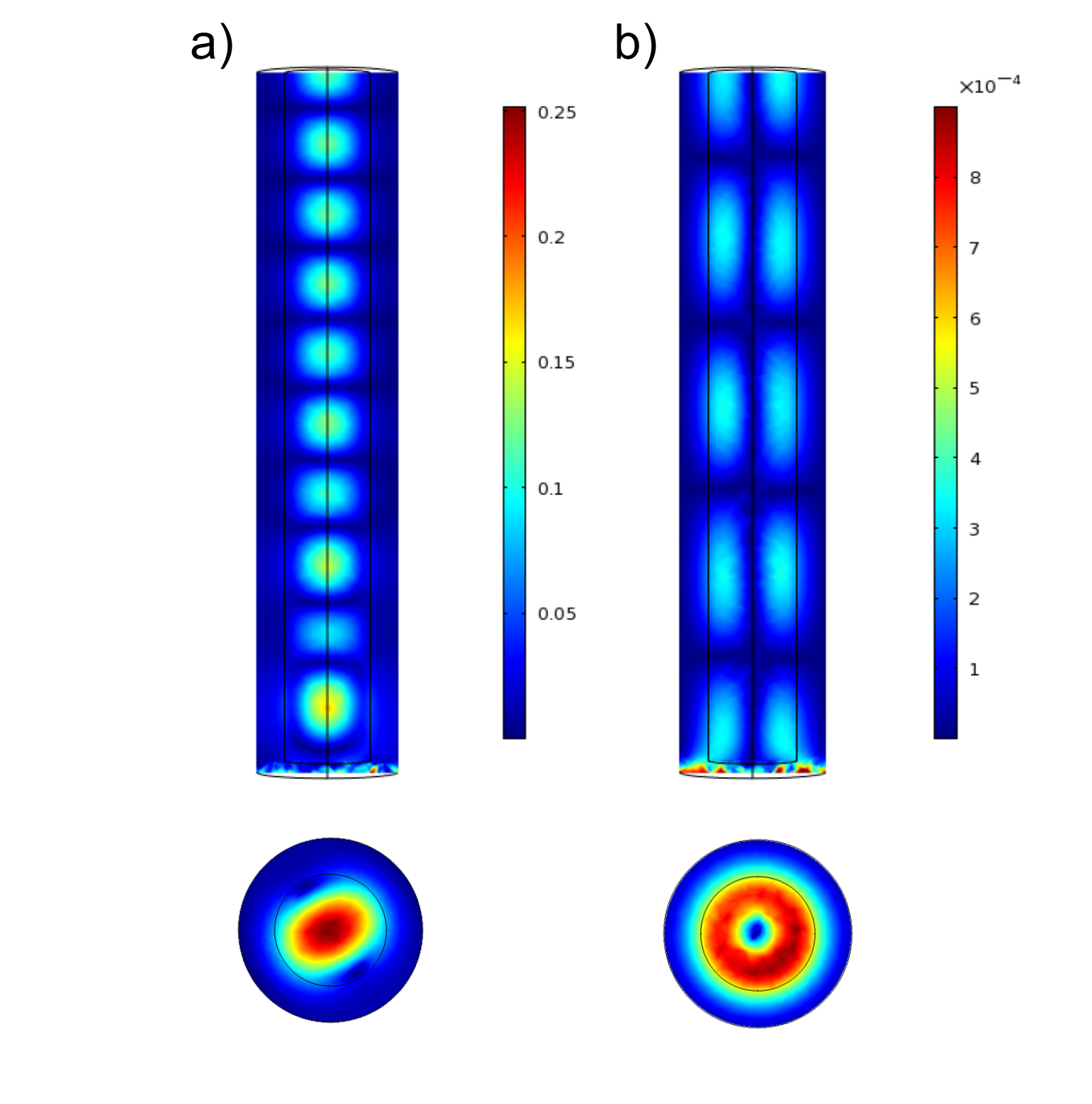}
\end{center}
\caption{$B_1$ map of traveling waves in a partially filled dielectric waveguide: a) TE$_{11}$ and b) TE$_{01}$, where top - coronal, bottom - axial slices (frequency domain). }
\label{3Dmodes}
\end{figure}
For simulation purposes, a metal cylinder was inserted inside the simulated magnet throughout the full bore length, thereby setting the metal boundaries of the waveguide. Within this metal cylinder, a tube of dielectric (deionized water with $\varepsilon_r = 80$ and $\sigma = 5 \times 10^{-5} S/m$) was fitted tightly such that propagating modes inside the dielectric would develop as pure modes of the cylindrical waveguide. Depending on the excitation method, the above setup with $D_{mesh}=5.6$ cm and dielectric $D_{rod}=3.4$ cm would support multiple modes up to a modified minimum critical wavelength $\lambda'_c $. 
Primary modes selection occurs due to the loop positioning with respect to the dielectric as described in Sec.\ref{theory}.  We also assumed that the loop coil was placed as close as possible to the end of the dielectric rod and centered it in the transverse plane. 
We computed the distribution of $B_1$ field of dominated modes in such waveguide by using 2D eigenmode solver as shown in Fig.~\ref{modes}. As follows from the field pattern, we confirmed that the waveguide support TE$_{01}$ mode (Fig.~\ref{modes}(a)) with the highest efficiency for the parallel geometry and the lowest TE$_{11}$ mode (Fig.~\ref{modes}(b)) that can be excited by the orthogonal coil geometry. The TE$_{01}$ field distribution in the transverse plane shows null of $B_1$, while $B_z$ is maximum at the center of the rod. The electric field (not shown in simulations) is directed along concentric circles around the null in the transverse plane of the rod. The TE$_{11}$ field distribution in the transverse plane shows maximum of $B_1$ in the center of the rod with the two nulls on the edges of the rod, where the $B_z$ components are maximum and aligned in opposite directions. The electric field of this mode is also maximum at the center of the rod and polarized perpendicular to $B_1$. As seen from the field pattern, only non-symmetric modes ({\em e.g.} TE$_{11}$) support circular polarization. All the fields evanescently decay outside the dielectric rod. 
Several TM modes (TM$_{01}$ and TM$_{11}$) are also supported by such system but with much smaller efficiency compared to the two primary TE modes so we do not show them here.

\begin{figure}[t!]
\begin{center}
\includegraphics[width=3.4 in]{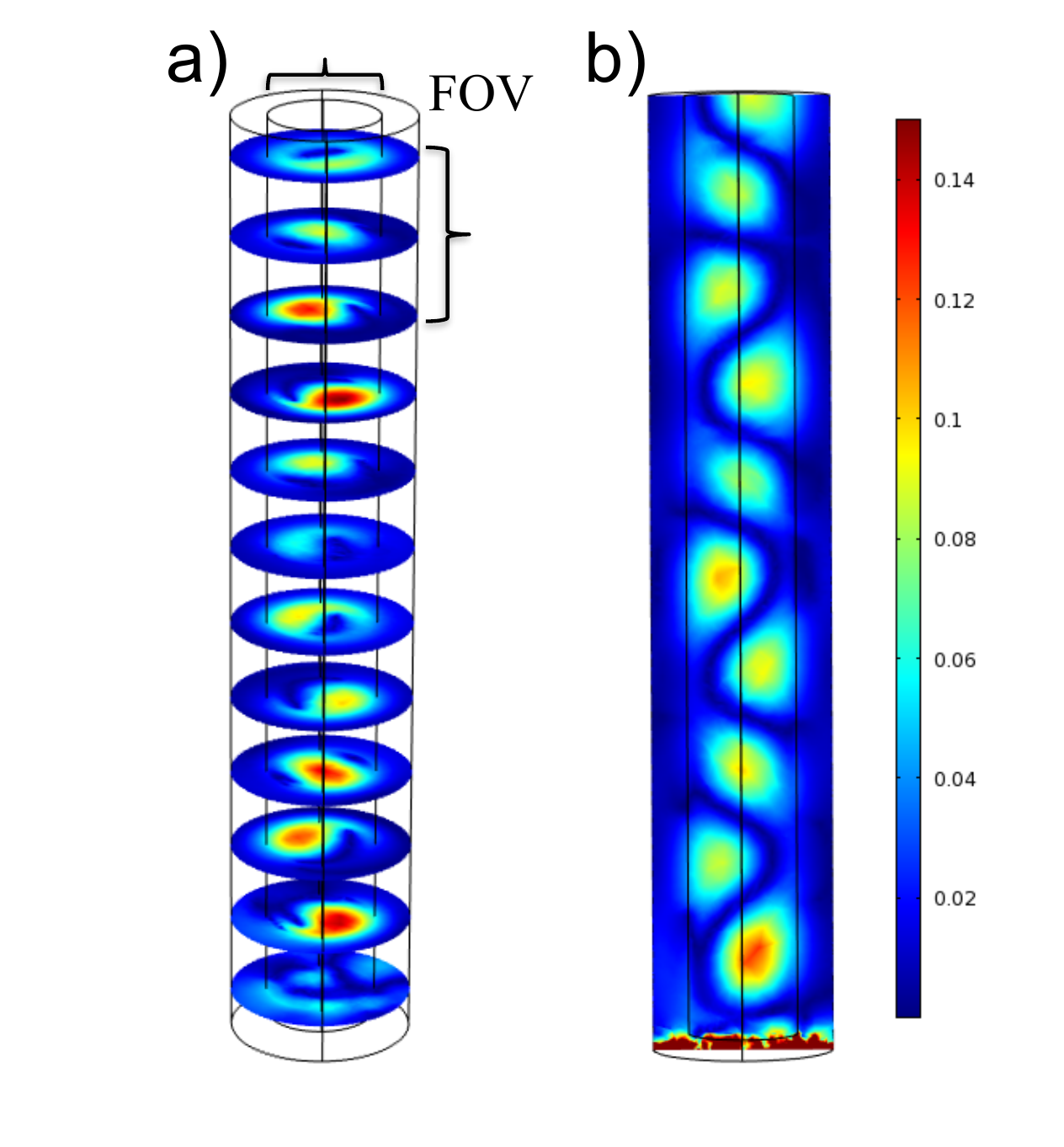}
\end{center}
\caption{ $B_1$ map of traveling wave in a partially filled dielectric waveguide: a) axial, b) coronal slices (frequency domain). }
\label{mix}
\end{figure}
We also simulated traveling wave propagation along entire waveguide in 3D using a frequency domain solver and a boundary mode analysis near 900 MHz with two numeric ports, one of which (in the bottom) has an RF power input. Depending on the number of allowed modes and eigenvalue set in the solver we observed various cases: near single mode and multimode (mixed) regimes. 

Figure~\ref{3Dmodes} shows coronal and axial slices of $B_1$ field map of traveling waves in a partially filled dielectric waveguide for single mode regimes with TE$_{11}$-like and TE$_{01}$-like modes.  
Simulation figures show a static picture of fringes in coronal plane that correspond to several periods of traveling waves. Such static picture is also equivalent (for field magnitude) to a standing wave pattern along {\em z}-axis. In practice, perfectly impedance matched waveguide results in washing out the fringes along {\em z}-axis. Since the TE$_{01}$ has smaller critical wavelength (that is harder to fulfill) than the TE$_{11}$ the period of TE$_{01}$ is approximately twice of the TE$_{11}$ period (see Fig.~\ref{3Dmodes}), so that $h_{11} > h_{01}$.
Note, that in the actual MR experiments only the top $6 \times 3.4$~cm (FOV) is imaged.

We also obtained a multimode regime where at least two modes propagate simultaneously along the guide. The typical field pattern of such regime is shown in Fig.~\ref{mix}.  The standing wave pattern in coronal plane along {\em z}-axis appears irregular with a spiral null pattern in 3D. In practice, multimode regime can be altered by insertion of additional dielectrics with different $\varepsilon_r$ \cite{BrunnerMRM2011} or conductors that effectively act as filters.  

\subsection{Experimental Method}
\label{experiment}
For experimental verification of the above simulations, the traveling wave setup using a cylindrical waveguide with a concentric dielectric sample was used on the 10.5-cm ultra-widebore, 21.1-T vertical magnet at the National High Magnetic Field Laboratory. The system is equipped with an Avance III 900 MHz spectrometer (Bruker BioSpin Corp, Billerica, MA) and a specially-built 6.35-cm I.D. microimaging (1 T/m/A peak) gradient set (Resonance Research, Inc., Billerica, MA). 
Simple loop coils were used in this study to excite modes within the cylindrical waveguide and receive MR signals for imaging. The coils were designed to fit into an existing wide-bore animal imaging probe frame. The remote transcieve loop coil was placed at the entrance to the cylindrical waveguide, with the outer border of the waveguide defined by a solid copper screen. Immediately adjacent to the loop coil, the dielectric rod was positioned concentric to the copper cylinder. The distance from the loop coil to the active imaging volume was approximately 25 cm. 
\begin{figure}[tb!]
\begin{center}
\includegraphics[width=3.5 in]{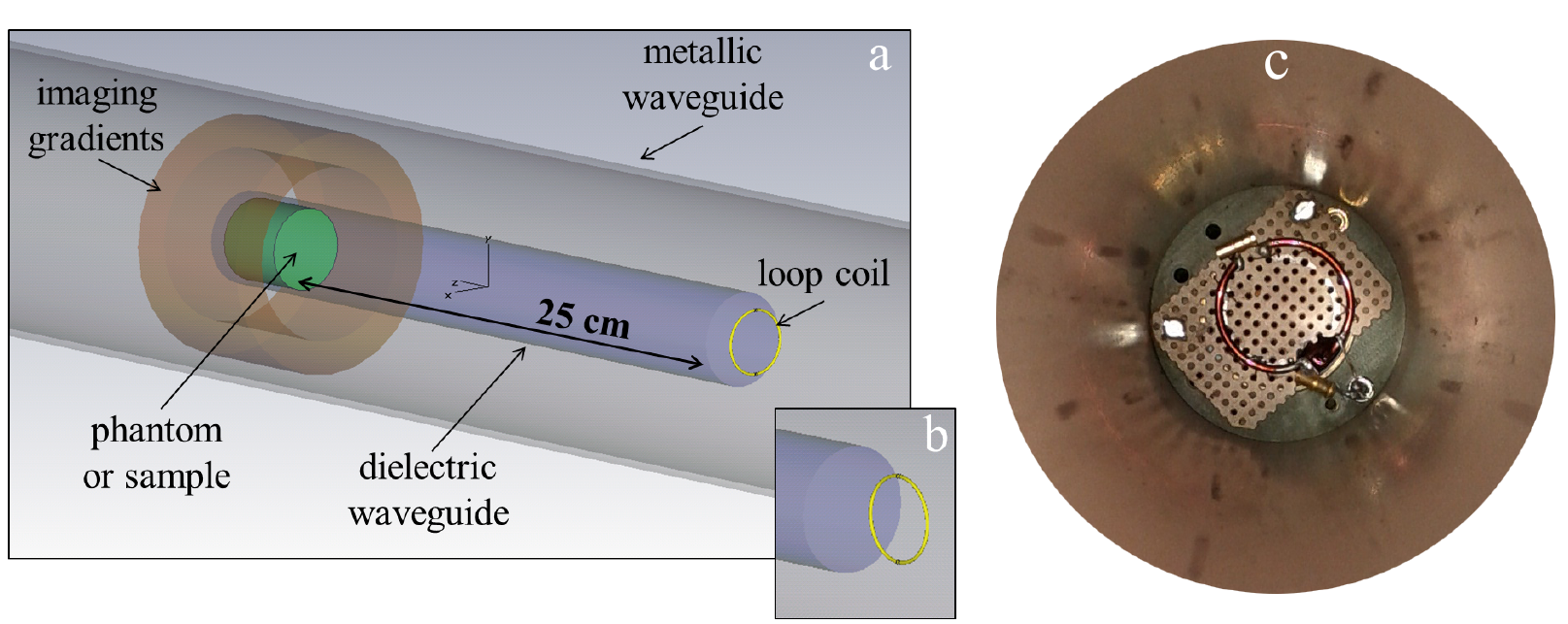}
\end{center}
\caption{Schematic of setup with parallel (a) and orthogonal (b) loop-coil geometries; a photograph of a loop-coil inside a metal guide (c). }
\label{setup}
\end{figure}
Figure~\ref{setup} displays the setup with the remote loop-coil (parallel or orthogonal to the transverse, 
{\em xy}-plane) and a waveguide inserted into the magnetÕs bore and imaging gradients.

The RF coil was constructed from 18-gauge copper wire resulting in a 2.4-cm diameter single loop. The coil was split once with a Giga-Trim non-magnetic variable capacitor (Johanson Manufacturing Corpoaration, Boonton, NJ) and was impedance matched to 50~$\Omega$ via an L-type network with similar variable capacitors (Fig.~\ref{setup}(c)). 
The copper screen (5.5-cm diameter, 29-cm length) was constructed from readily available copper shim stock such that it fit onto an existing in vivo animal probe and inside the imaging gradients. Figure~\ref{setup}(c) shows the part of actual probe body with the separate waveguide mounted on the probe frameÕs lower flange at the same location where the loop coil is located.

Most of the experiments were performed within the 5.5-cm copper waveguide and the loop coil placed approximately 25-cm away from the imaging volume. In a reference experiment we placed a deionized (DI) water phantom in a 50-mL centrifuge tube at the imaging volume. We observed no signal because cut-off requirements were not met with an open air waveguide and the coil itself is not operated in the far-field regime. With the introduction of a high permittivity dielectric into the waveguide between the loop coil and imaging volume, the critical wavelength is increased, and several propagation modes can be supported. To image these modes inside the dielectric guide itself at the gradient insert region, data were acquired using a 2D FLASH sequence with TE/TR = 5.0/500 ms, $8 \times 4$-cm FOV and 312-$\mu$m in-plane resolution and a 3D FLASH sequence with TE/TR = 5.0/150 ms, $6.5 \times 3.8 \times 3.8$-cm FOV and 100-$\mu$m isotropic resolution. Transmission power was set by maximizing the receive signal. 

In all instances (except where noted), the dielectric rod used to propagate the traveling waves was also the sample under investigation. For all experiments, DI water was used as the main high permittivity ($\varepsilon_r \approx 80$) dielectric. We also varied dielectric rod diameters to 2.4, 3.4 and 4.7 cm in separate experiments to evaluate the impact of diameter of dielectric to the propagating mode. In all cases, the length of the tube extended to the end of the waveguide. For $D_{rod}=3.4$-cm, tests were conducted with the loop coil in orientations parallel and orthogonal to the $B_0$ field. Furthermore, experiments were conducted with and without the presence of the copper screen to utilize metal bore of the spectrometer.

\section{Results and Discussion}
\label{result}
With the TW NMR setup shown here, images from multiple samples were successfully acquired with a loop coil placed 25-cm away, covering a FOV that is almost twice the coilÕs linear dimensions in the transverse plane and approximately 35 times in the longitudinal plane (limited by the gradientÕs physical dimensions). Despite certain imaging artifacts and nulls inherent to TW modes here, one of the main benefits of TW NMR is the ability to remotely image entire volumes that provide significantly more coverage than a birdcage coil, a feature that is limited by the usable length of the systemÕs imaging gradients. 
\begin{figure}[h!]
\begin{center}
\includegraphics[width=3.2 in]{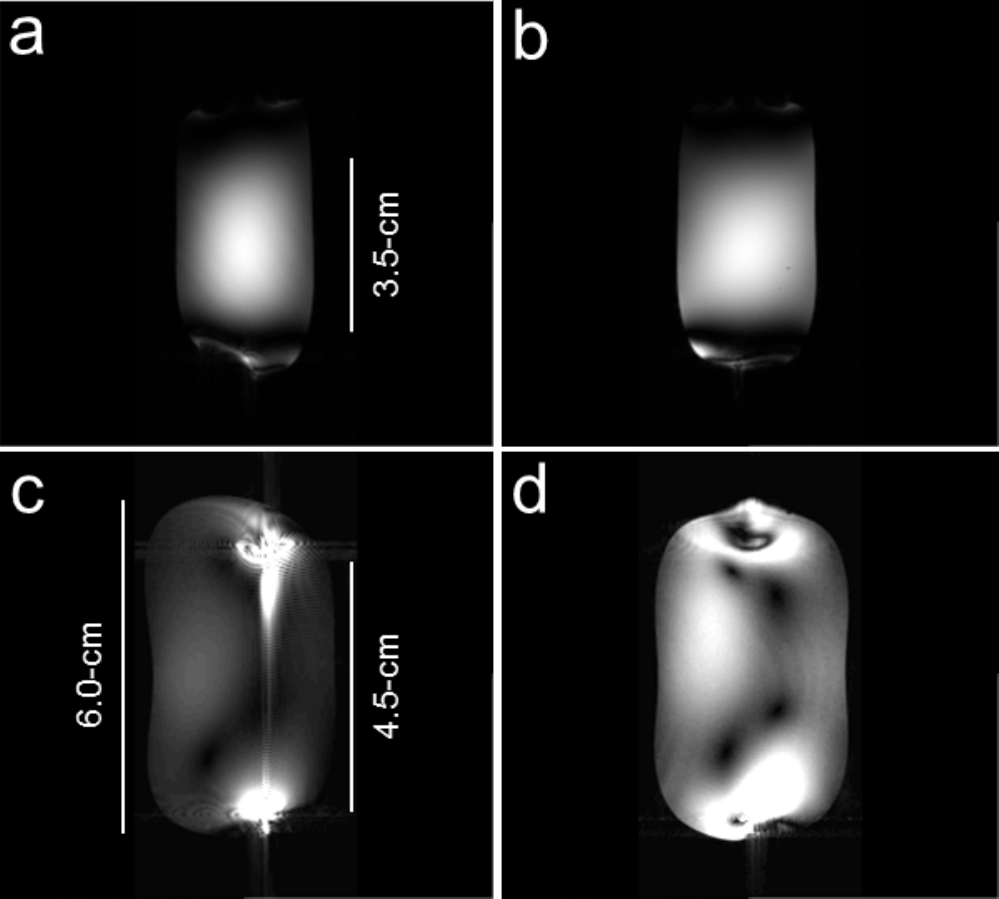}
\end{center}
\caption{Images obtained from liquid samples placed within a birdcage coil (a, b) and within the TW setup using the parallel loop coil (c, d). }
\label{bc}
\end{figure}

Figure~\ref{bc}(a,~b) compares images from sagittal and coronal slices obtained from liquid samples placed within a birdcage coil and within the TW setup using the parallel loop coil (Fig.~\ref{bc}(c, d)). Notice the birdcage coil covers a length of 3.5-cm and the TW setup covers as far as the gradients can ``see'', a length of about 6.0-cm. The warping and brightening that occur at the edges of the images are indicative of the gradient edges that result in encoding errors. It should be noted that the samples have different diameters limited by the each coilÕs setup: 2.5-cm for the birdcage coil and 3.5-cm for the traveling wave setup.

For the liquid samples imaged in this work, two main issues were consistent throughout the experiments. One is a hyper-intense zipper artifact that is always present only at the center slice of the imaging FOV and does not appear in slices adjacent to the center slice. The artifact appears as two parallel zippers on opposite ends of the image edges (see Figs.~\ref{bc} (c, d)). 
There also exists another single zipper artifact running perpendicular to the other two zippers, which are located mainly in the longitudinal slices (sagittal and coronal), whereas axial images only ``see" the artifact as a few bright pixels. These artifacts are common in spectrometers and attributed to the geometric limits of gradient insert and associated encoding error.

The other observed imaging issue with the setup is the signal inhomogeneities caused by the nulls that are inherent to specific modes in a single or multimode regimes within the waveguide. Interestingly, the nulls proved useful when comparing the different waveguide setups mentioned earlier. While the nulls appeared to have similar patterns, the main difference between the different waveguide setups was the nullsÕ description of the wave period and phase. The same can be said about experiments with different RF loading or samples. 
\begin{figure}[h!]
\begin{center}
\includegraphics[width=3.2 in]{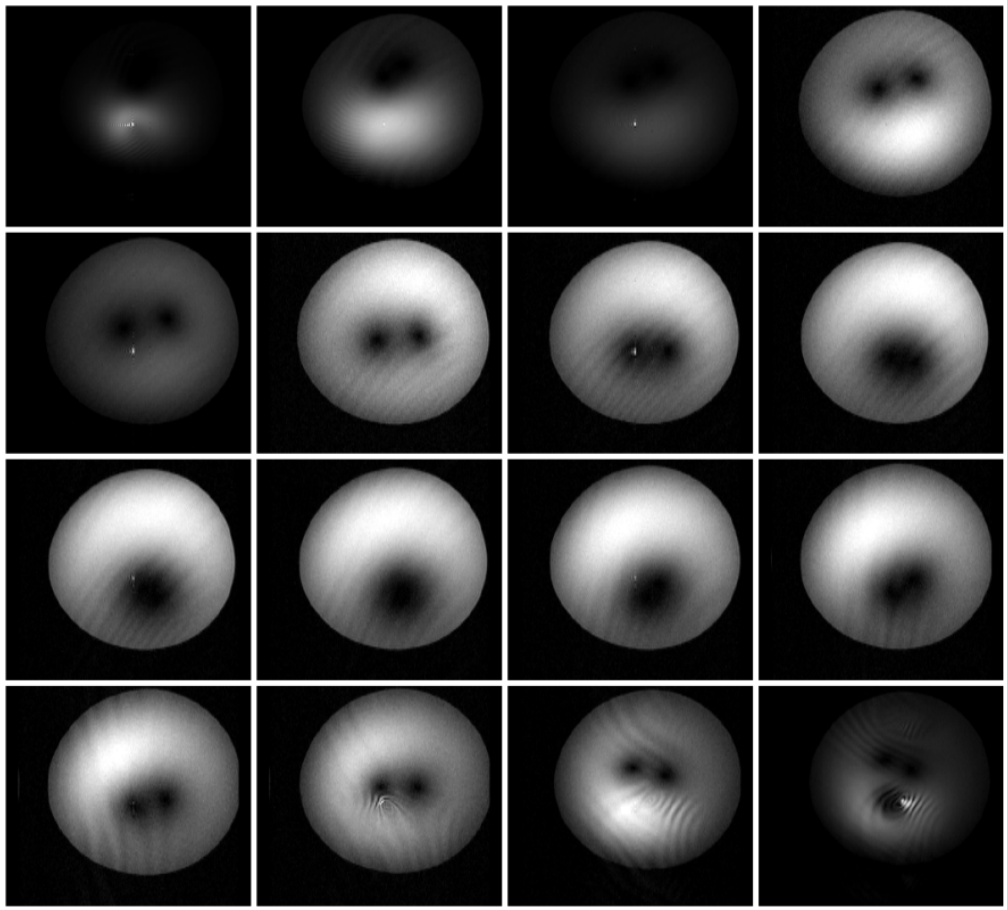}
\end{center}
\caption{Array of axial slices ($D=3.4$~cm) that begin at the bottom and end at the top of the FOV for the DI water sample utilizing the parallel coil setup.}
\label{parcoil}
\end{figure}

Figure~\ref{parcoil} represents an array of axial slices that begin at the bottom and end at the top of the FOV within the DI water sample using the parallel loop-coil, dielectric rod of $D_{rod}=3.4$-cm, and separate 5.5-cm copper waveguide setup (this setup is similar to the one that used to obtain Figs.~\ref{bc}(c, d). 
It appears as if the two nulls are spiraling away from each other as the mode propagates along {\em z}-axis. This behavior is evident for most of the samples with the parallel coil setup, differing in period and phase. The separate dots (nulls) in the axial slices correspond to the propagating and partially reflected waves. For small flip angle and in the far rf field regime approximation the image pattern is commensurable to $B_1$ field map. So, apart from the reflected wave (second null) the observed pattern shows TE$_{01}$-like mode, which is slightly offset from the center. TE$_{01}$-like mode in the parallel case in Fig.~\ref{parcoil} matches respective simulated mode in 
Fig.~\ref{modes}(a). 
\begin{figure}[tb!]
\begin{center}
\includegraphics[width=3.2 in]{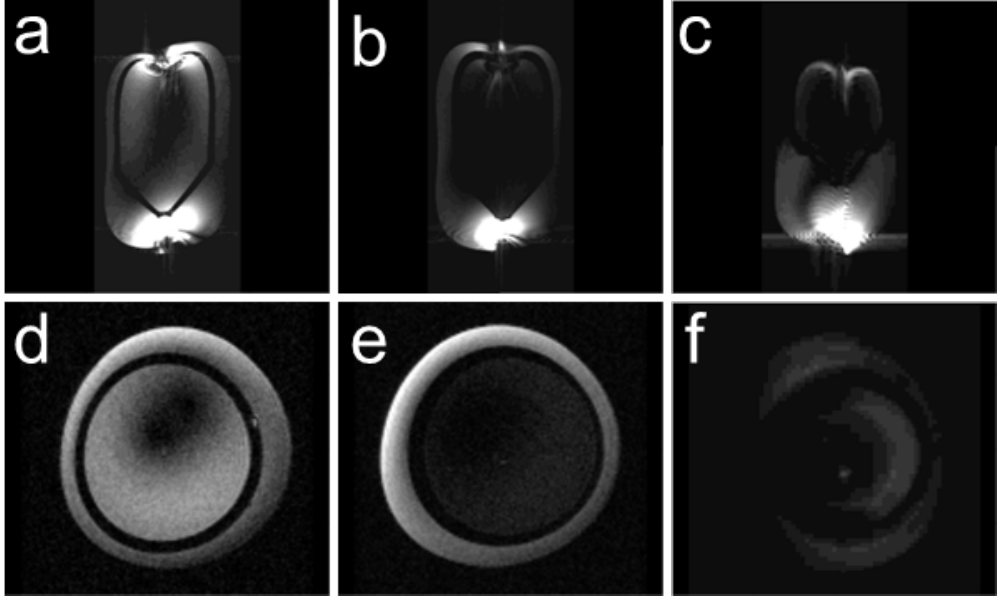}
\end{center}
\caption{Central coronal (a-c) and axial (d-f) slices of the $1\times$PBS (a, d)), PEG (b, e), and 2.5 M NaCl (c, f) samples, respectively.}
\label{pbs}
\end{figure}
%
\begin{figure}[tb!]
\begin{center}
\includegraphics[width=3.4 in]{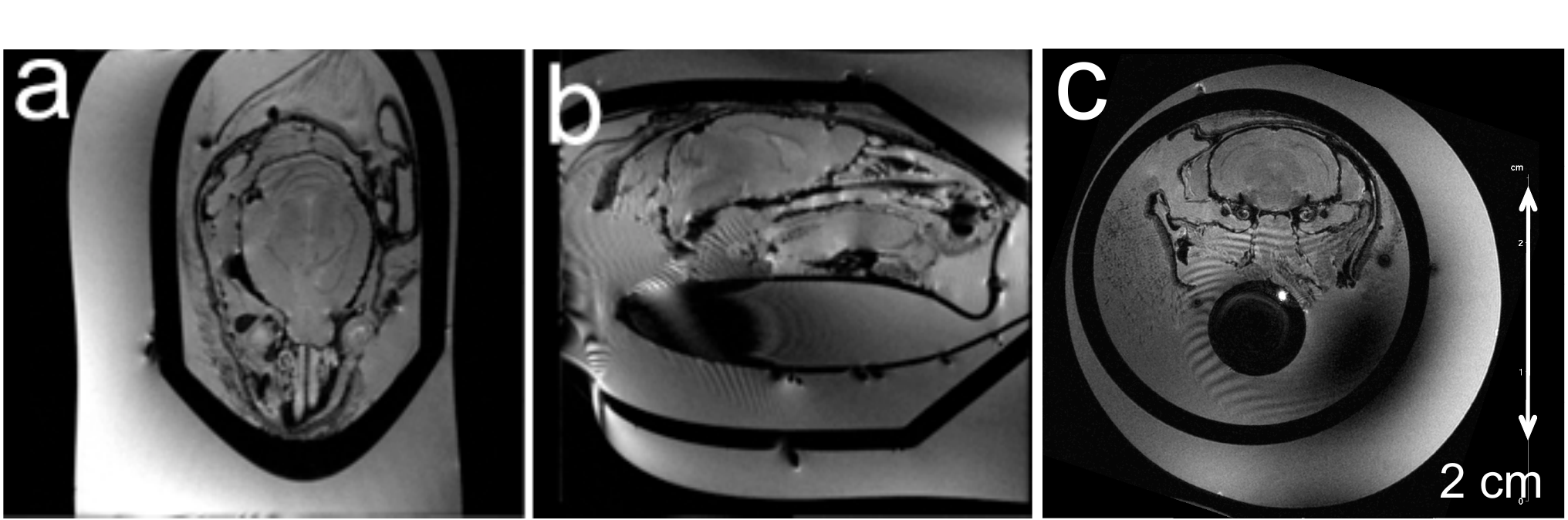}
\end{center}
\caption{{\em Ex-vivo} mouse head immersed in $1\times$PBS and placed within the 50-mL centrifuge tube: a) coronal, b) sagittal, c) axial slices.}
\label{mouse}
\end{figure}
%

Using parallel coil setup we imaged various dielectric samples placed into centrifuge tubes, which were inserted inside dielectric guide within FOV provided by the gradient insert. Figures~\ref{pbs}(a-c) show a coronal central slice of the $1\times$PBS, PEG, and 2.5 M NaCl samples, respectively. Figures~\ref{pbs}(d-f) show a central axial slice of the same samples. While the null patterns are similar to those seen with the DI water samples (see Figs.~\ref{bc}(c-d), and Fig.~\ref{parcoil}), the efficiency at which mode propagates into the sample varies due to different conductive and dielectric properties of the samples. The highest SNR is obtained for $1\times$PBS with $\varepsilon \sim 80$ (see Figs.~\ref{pbs}(a, d)) and the lowest SNR -- for PEG (400) with $\varepsilon = 12.4$ (see Figs.~\ref{pbs}(b, e)). The NaCl sample (see Figs.~\ref{pbs}(c, f)) shows nonuniform distribution along {\em z}-axis, in agreement with the high conductivity that damps the wave propagation.

The feasibility of tissue imaging with the setup used here is shown in Fig.~\ref{mouse}, which is a central result of our work. The fixed C57 mouse head was immersed in $1\times$PBS and placed within the 50-mL centrifuge tube. Despite the persistent signal nulls inherent to TE$_{01}$, the setup here was able to image {\em ex-vivo} mouse head with decent homogeneity within the brain tissue area. As can be seen in Fig.~\ref{mouse}(c), the null entering the sample side appears to be located, and remained on the edge opposite of the actual mouse brain. 

\begin{figure}[tb!]
\begin{center}
\includegraphics[width=3.4 in]{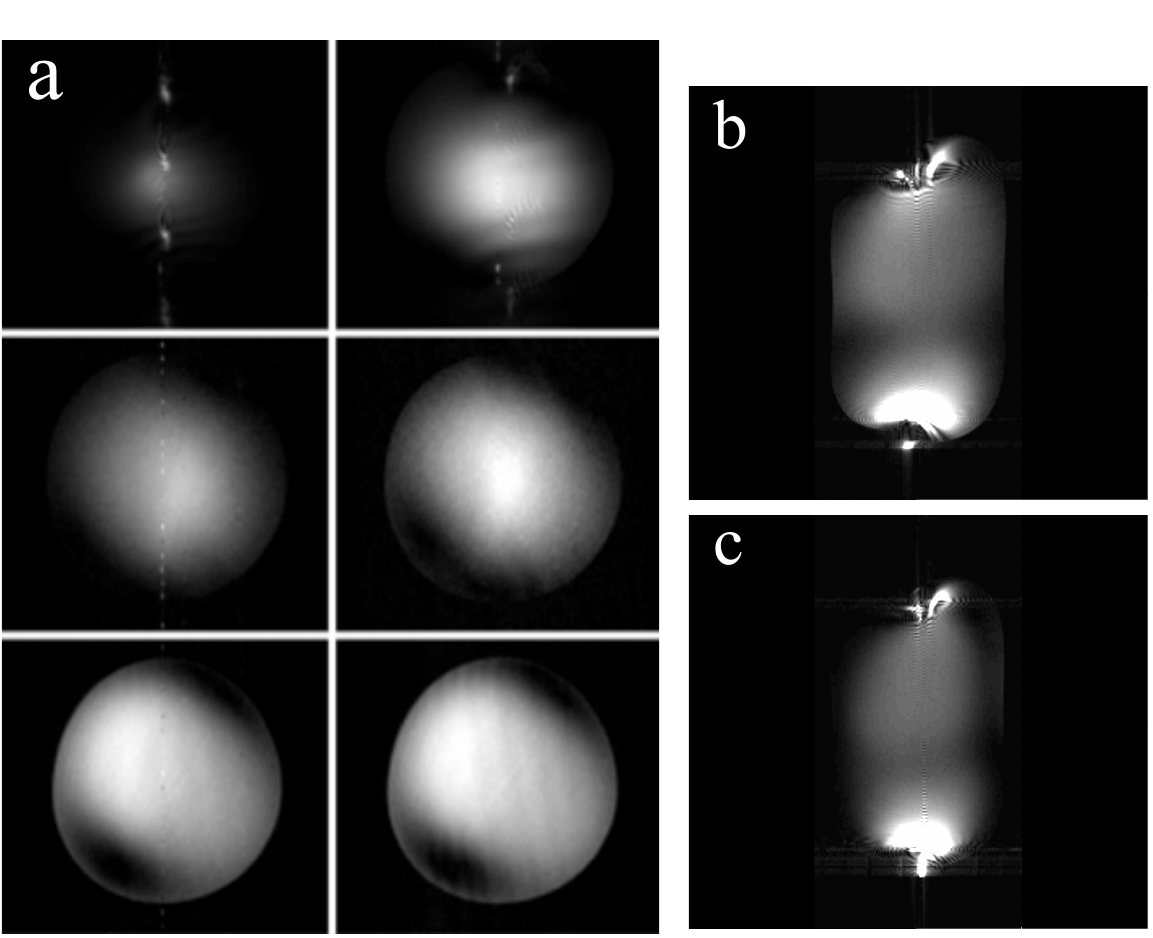}
\end{center}
\caption{Array of axial slices ($D=3.4$~cm) that begin at the bottom and end at the center of the FOV a); coronal b), and sagittal c) slices for the DI water sample utilizing the orthogonal coil setup.}
\label{orth}
\end{figure}
We repeated experiments with a same setup as per results shown in Fig.~\ref{parcoil} with the similar coil, which was oriented in orthogonal plane. Figure~\ref{orth} summarizes axial images of DI water utilizing the orthogonal coil setup. For properly matched coil, as expected from the theory and simulations, the null patterns differ from the ones obtained in the parallel coil case. In particular, TE$_{11}$-like mode in the orthogonal case in Fig.~\ref{orth} matches respective simulated mode in Fig.~\ref{modes}(b). The orthogonal coil setup seems more suitable for uniform excitation of the sample placed in the center of FOV, while parallel setup provides excitation of the peripheral areas. The SNR analysis  confirms the preference of orthogonal coil arrangement giving maximum SNR=92 for the central axial slice in Fig.~\ref{orth}~(a), as compared to the SNR=79.2 of a similar slice for the parallel coil arrangement (Fig.~\ref{parcoil}). For the reference we also calculated SNR for the birdcage coil and PEG sample to be 156 (Fig.~\ref{bc}~(a)), whereas TW image of PEG gives 4.85 (Fig.~\ref{pbs}~(e)).

For all the above presented experimental data the field patterns were consistent for fixed geometry and coil tuning, however, we observed fluctuation of the coil tuning with respect to slight offset of the coil and dielectric, which was inherent to different experimental runs. Such tuning instabilities generally resulted in multimode regime and ``contaminated" mode patterns for either coil geometry. From engineering point of view this problem can be fixed by designing a special probe with built in loop-coil and dielectric guide. 
We also designed a special mode filter made of multiple dielectric tubes, which were arranged in a star-like (triangular lattice) pattern. Such anisotropic mode filter placed in the bottom of the FOV helped to clean the propagating mode by affecting all non-symmetric modes and allowing perfect TE$_{01}$ mode in the parallel coil geometry. In a similar fashion, the manipulation of any modes can be done by careful arrangement of the dielectric tubes at the field extrema. 

We further verified our screened dielectric concept with various screens and dielectrics arrangements. Specifically, we removed a copper shield and placed the same dielectric tube (D=3.4 cm) inside the scanner thus changing the diameter of the metal screen to a maximum available bore diameter (D=10.5 cm) that stepped down to about 6 cm inside the gradient insert. We were able to couple same modes into our waveguide and obtained MR images although with decreased SNR that we attribute to small effective permittivity~\cite{Wu2001} in such low volume fraction guide. We also varied diameters of the dielectric tube with the fixed metal screen diameter, {\em e.g.} increased guide fill factor. For the dielectric diameter of 4.7 cm we obtained multimode pattern in both coil configuration with higher propagation constant as observed from the partial standing wave pattern. The above observations qualitatively agree with the theory. While dielectric insert with large diameter allows more propagating modes, the smaller diameter may be preferred for a single mode regime.  

\section{Conclusion}
Traveling wave MRI in an ultra-high field vertical MR system was achieved with an appropriate cylindrical waveguide and a high permittivity dielectric sample to enable propagation beyond hollow guide cut-off for imaging in the far field. Theory and simulations corroborate the observed propagating mode structure. The applicability of the above methods relies on propagating wave properties, which are much more pronounced at ultra-high fields (due to smaller wavelengths) of small bore systems. While these systems potentially benefit most from this implementation, the impact and utilization of high dielectric materials/samples as an integral part of the waveguide may be significant for clinical MRI systems, particularly with continuing increases in $B_0$ field strength that are under construction and in planning. Furthermore, the SNR in presented  results can be improved with a more efficient guide excitation (index matching) by incorporation of the coil inside the dielectric. 

Our results show a simple way to implement multimode regime for potential parallel imaging schemes using only a single (or a few) channel. The fundamental implication of this work is the exploration of new phenomena arising from the spatial and spectral interference of propagating excitation fields. 
Applicability of the above method rely on propagating waves properties, which are much more pronounced at the ultra high fields (due to smaller free-space wavelengths) of small bore scanners (spectrometers), therefore, those systems potentially benefit even more from traveling waves than human size systems.

This work was funded by NSF (DMR-0084173 and NHMFL User Collaborations Grant Program).
%
%

\bibliographystyle{model1-num-names}
\bibliography{NMRreferences}

\end{document}